\documentclass[letterpaper,journal]{IEEEtran}
\usepackage[utf8]{inputenc}
\usepackage{textcomp}
\DeclareUnicodeCharacter{2212}{\textminus}
\usepackage{amsmath,amsfonts}
\usepackage{algorithmic}
\usepackage{algorithm}
\usepackage{array}
\usepackage[caption=false,font=normalsize,labelfont=sf,textfont=sf]{subfig}
\usepackage{textcomp}
\usepackage{stfloats}
\usepackage{url}
\usepackage{verbatim}
\usepackage{graphicx}
\usepackage{booktabs}
\usepackage{cite}
\usepackage{amsmath,amssymb}
\makeatletter
\newsavebox\myboxA
\newsavebox\myboxB
\newlength\mylenA

\usepackage{xcolor}
\newcommand*{\corr}[1]{\textcolor{black}{#1}}
\newcommand*\xoverline[2][0.75]{%
    \sbox{\myboxA}{$\m@th#2$}%
    \setbox\myboxB\null
    \ht\myboxB=\ht\myboxA%
    \dp\myboxB=\dp\myboxA%
    \wd\myboxB=#1\wd\myboxA
    \sbox\myboxB{$\m@th\overline{\copy\myboxB}$}
    \setlength\mylenA{\the\wd\myboxA}
    \addtolength\mylenA{-\the\wd\myboxB}%
    \ifdim\wd\myboxB<\wd\myboxA%
       \rlap{\hskip 0.5\mylenA\usebox\myboxB}{\usebox\myboxA}%
    \else
        \hskip -0.5\mylenA\rlap{\usebox\myboxA}{\hskip 0.5\mylenA\usebox\myboxB}%
    \fi}
\makeatother

\begin{document}

\title{3D-Structured Polyethylene Windows for Low-Loss Transmission in Wideband Cryogenic Terahertz Systems}

\author{Fran\c{c}ois Joint, Igor Lapkin, Pierre-Baptiste Vigneron, Emilie Hérault, Denis Meledin, Alexei Pavolotsky, Magnus Strandberg, Sven Erik Ferm, Mathias Fredrixon, Leif Helldner, Erik Sundin, Victor Belitsky, Vincent Desmaris
\thanks{F. Joint, I. Lapkin, D. Meledin, A. Pavolotsky, M. Strandberg, S-E .Ferm, M. Fredrixon, L. Helldner, E. Sundin, V. Belitsky, V. Desmaris are with the Group for Advanced Receiver Development, Onsala Space Observatory, Department of Space, Earth and Environmental Sciences, Chalmers University of Technology, Gothenburg, SE-41296, Sweden}
\thanks{P-B Vigneron and E. Hérault are with Université Grenoble Alpes, Université Savoie Mont Blanc, CNRS, Grenoble INP, CROMA, Grenoble, 38000, France}
}



\maketitle

\begin{abstract}
We present the design, fabrication, and characterisation of a broadband vacuum window and infrared filter based on ultra-high molecular weight polyethylene (UHMWPE) for millimeter wave receivers operating across ALMA Band 6 and 7 (211–373 GHz). The window incorporates pyramidal anti-reflection (AR) structures, machined directly into the polyethylene using CNC machining, which \corr{provide} impedance matching over a broad frequency range. The structured UHMWPE method was implemented in two distinct components: a vacuum window and a cryogenic infrared filter. This surface-structuring approach provides mechanical robustness, cryogenic compatibility, and low insertion loss. We characterize the transmission properties using Terahertz Time-Domain Spectroscopy (THz-TDS), which demonstrates reflection below $5\%$ across the full band. Complementary heterodyne measurements confirm improved receiver noise performance. These results establish 3D-structured UHMWPE as a promising platform for broadband cryogenic optics in high-sensitivity THz instrumentation.
\end{abstract}

\begin{IEEEkeywords}
Atacama Large Millimeter/submillimeter Array, Vacuum window, Broadband anti-reflection structures, Heterodyne noise temperature. 
\end{IEEEkeywords}

\section{Introduction}
\IEEEPARstart{C}{ryogenic} vacuum windows and infrared filters serve distinct yet complementary roles in millimeter and sub-millimetre wave receiver systems for radio astronomy. Vacuum windows must maintain pressure integrity while offering high transmission and minimal reflection over a broad spectral range. Infrared filters, usually placed at cryogenic stages, are designed to block thermal radiation from warmer optics while maintaining low insertion loss at the signal frequency. Both components are critical to preserving receiver sensitivity and minimizing unwanted optical loading. In advanced observatories like ALMA, where high-frequency heterodyne receivers operate across wide bandwidths, even minor transmission losses at the window or IR filter interface can have a significant impact on overall noise performance \cite{yagoubov_wideband_2020, speirs_design_2020, schroder_design_2016}.

Dielectric windows made of materials such as high density polyethylene (HDPE) or quartz are commonly used due to their low loss tangent and good mechanical properties. However, these materials typically exhibit large refractive index mismatches with air if no anti-reflection strategies are adopted, leading to high Fresnel reflections for each interface. This reduces the transmission and also introduces standing waves and coupling artifacts that increase the overall receiver noise temperature.

Among available plastic materials for window manufacturing, ultra-high-molecular-weight polyethylene (UHMWPE) offers a compelling combination of mechanical robustness and low-loss optics \cite{yagoubov_wideband_2020}. Mechanically, UHMWPE exhibits a tensile strength of $\approx$ 40 MPa, twice that of HDPE, allowing thinner windows that deform less under a 1 bar pressure differential \cite{dalessandro_ultra_2018}. \corr{For a 10 mm-thick UHMWPE sheet, the transmission is $\geq$ 90 $\%$ from 150-800 GHz, with refractive index 1.537 and absorption coefficient of 0.03 $Np\cdot cm^{-1}$ at 300 GHz}~\cite{dalessandro_ultra_2018}. \corr{For cryogenic IR filtering, UHMWPE’s vibrational spectrum produces strong absorption in the mid‑IR: a 0.5 mm plate shows $\approx$40–70 $\%$ transmission (\textit{T}) across 8–12 $\mu m$ with a peak  $\approx$77$\%$ at 10.5 $\mu m$ under optimized processing, while 2 mm becomes effectively opaque in 8–12 $\mu m$ ($T\leq20\%$) \cite{liu_obtaining_2022}.}

To overcome Fresnel reflections, various anti-reflection (AR) strategies have been proposed. Conventional thin-film quarter-wave coatings offer narrowband solutions but suffer from limited thermal compatibility and poor adhesion under cryogenic cycling \cite{joint_compact_2019,joint_terahertz_2024,chen_anti-reflection_2014}. Multilayer dielectric coatings can achieve broader bandwidths and improved impedance matching, as demonstrated for cryogenic aluminum oxide optics \cite{nadolski_broadband_2020}, but their fabrication complexity and sensitivity to thermal contraction often limit scalability or long-term durability. Metallic impedance-matching layers can further extend bandwidth but introduce significant absorption losses incompatible with low-noise receivers. More recent efforts have focused on engineered metamaterials and sub-wavelength structured surfaces such as “moth-eye” textures and pyramidal grooves  which produce a graded effective refractive index profile to minimize reflections across a wide frequency range while maintaining mechanical robustness at cryogenic temperatures \cite{datta_large-aperture_2013, matsumura_millimeter-wave_2016, defrance_flat_2025}.

Among these, triangular or pyramidal relief structures have shown particular promise for broadband applications in the THz regime \cite{schroder_design_2016}. Their geometry enables a smooth impedance transition from air to dielectric, and their performance can be optimized for low polarization sensitivity and wide incident angle tolerance\cite{tapia_systematic_2018}. Fabrication approaches such as mechanical machining, lithographic patterning, and deep reactive ion etching (DRIE) have been employed to realize these structures on substrates ranging from polymers to high-resistivity silicon.

\corr{Despite this progress, we are not aware of demonstrations that design and experimentally validate a vacuum window over a continuous 211-373 GHz band ($\approx$55$\%$ fractional bandwidth). In this work, we present the design, fabrication and measurement of a pyramidal-textured UHMWPE window operating across this broadband range}. \corr{Since our target bandwidth is very broad ($\approx55\%$ fractional), the standard quarter‑wave formulas for 1D triangular grooves are not adequate. We therefore adopted a full‑wave FEM on a periodic unit cell with Floquet ports and an effective medium (EMT) model for rapid prototyping. The EMT model is verified with FEM, and both methods guided our choice of apex angle and pitch under machining constraints (see Sec.~\ref{Methods} and Fig.~\ref{fig_3d})}. The window is designed to meet the dual requirements of broadband transmission and mechanical robustness for cryogenic receiver systems. Optical performance was assessed using terahertz time-domain spectroscopy (THz-TDS), while system-level validation was performed in a heterodyne receiver setup.

Our results demonstrate broadband optical transmission between 95$\%$ and 99$\%$ across the 211–373 GHz range, with low insertion loss confirmed through both THz-TDS and heterodyne noise measurements. Although absolute reflection was not directly measured, the observed system-level performance indicates effective suppression of reflection-related losses. The proposed approach uses an established gradient-index AR principles applied to UHMWPE, implemented through scalable CNC machining of subwavelength pyramidal structures. This offers a practical, cryogenically compatible solution for broadband vacuum interfaces in millimeter-wave and terahertz receiver systems.

\section{Methods}
\label{Methods}
\subsection{Finite-Element Method}

\begin{figure}[!t]
\centering
\includegraphics[width=3in]{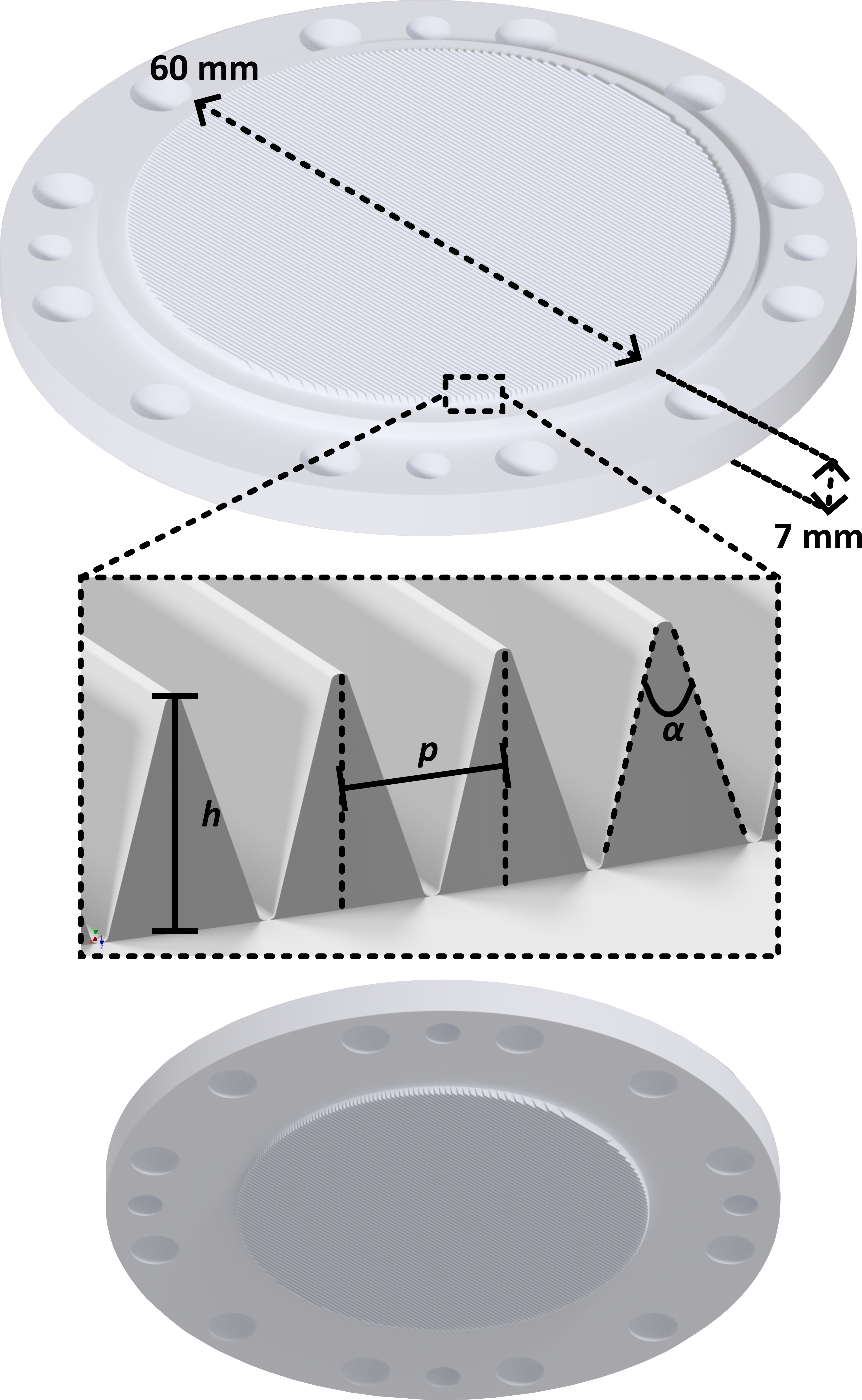}
\caption{Illustration of the UHMWPE vacuum window prototype.
}
\label{schematic}
\end{figure}

The electromagnetic response of the structured vacuum window as shown in Fig. \ref{schematic} and different types of AR profiles were analyzed using a full-wave solver \corr{(Ansys HFSS)} based on the finite element method (FEM). To exploit the periodicity of the structures, the window was modeled as an infinite two-dimensional periodic array and simulated using a single unit cell with periodic boundary conditions.
Excitation was provided by Floquet ports, which launch and receive orthogonal TE and TM modes \corr{We define the co‑polar component as the field parallel to the incident linear polarization, and the cross‑polar component as the orthogonal field. We also refer to TE when the incident E‑field is aligned with the AR grooves and to TM when it is orthogonal.} A frequency sweep from 211~GHz to 373~GHz (\corr{center frequency 292~GHz, 0.78–1.38~$\lambda_0$} bandwidth) was used, with mesh refinement near the pyramid apex and foundation to ensure convergence of return-loss below –20 dB across the band.

\begin{figure}[!t]
\centering
\includegraphics[width=3in]{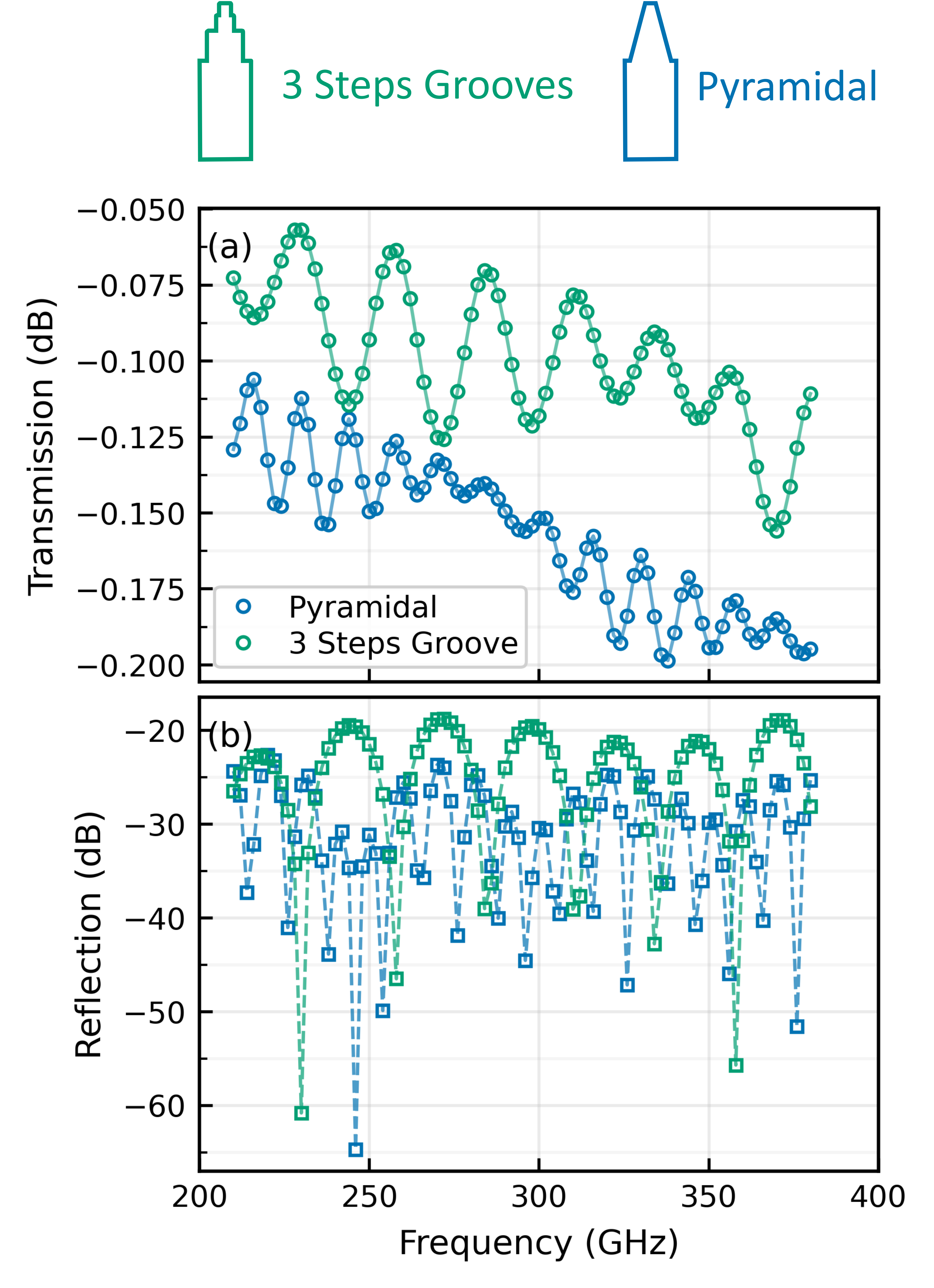}
\caption{\corr{Co‑polar performance of two broadband AR unit cells at normal incidence. 
(a) Transmission Tx and (b) reflection Rx for a pyramidal profile  and a three‑step groove profile. Curves for both incident linear polarizations (TE: E$\parallel$grooves,TM: E$\perp$grooves) are overlaid; co‑polar transmission/reflection are indistinguishable at normal incidence. Both unit cells share the same pitch (p) and depth (h); the three‑step layer heights/widths were chosen to approximate the triangular effective‑index gradient.}}
\label{fig_3b}
\end{figure}

\corr{We first evaluated two anti‑reflection (AR) unit cells: a continuously graded pyramidal profile and a three‑step groove profile. FEM simulations in Fig.~\ref{fig_3b} at normal incidence show that the three‑step unit cell yields only a small improvement in co‑polar transmission relative to the continuous pyramid across the band. Cross‑polar terms in both cases remained more than 20 dB below co‑polar over 211–373 GHz. The triangular profile also exhibits reduced in-band ripple and improved band-edge return loss compared with a stepped groove of same depth, optimized at mid-band. We interpret this with the gradient index approach: the triangular profile gives a smooth impedance transition, while a straight groove behaves as a quasi quarter-wave transformer with abrupt interfaces. Across 211–373 GHz, the FEM co‑polar transmission and reflection for TE and TM are indistinguishable at normal incidence for both design. The cross‑polar terms remain at the solver noise floor across the band. On this basis, we retained the continuous pyramidal profile as our baseline design.}

\begin{figure}[!t]
\centering
\includegraphics[width=3in]{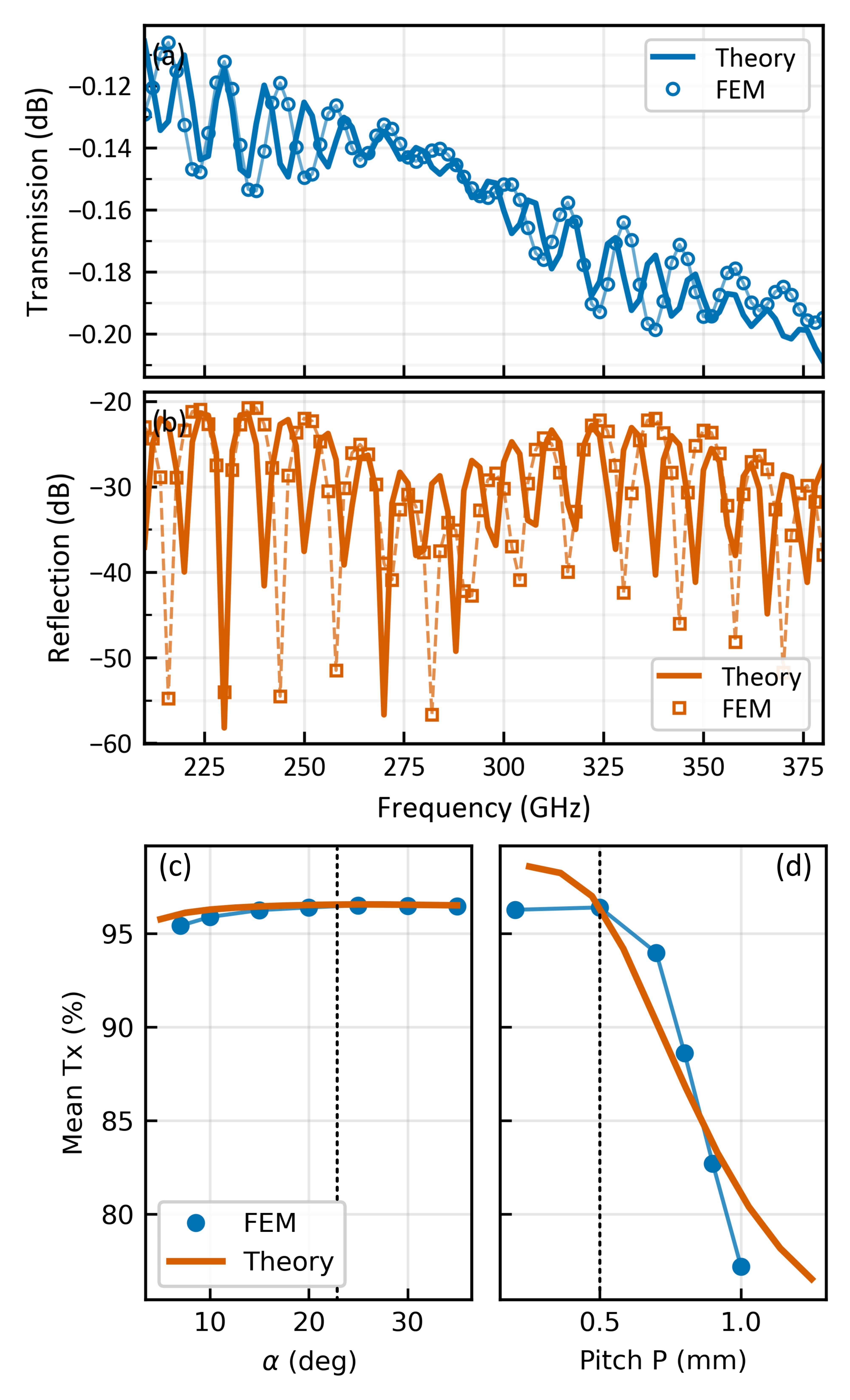}
\caption{\corr{Comparison of full‑wave (FEM) and effective‑medium (EMT) models for the UHMWPE pyramidal AR window and parametric sensitivity of the mean transmission. TE and TM polarisation are perfectly overlaid. (a) Normal‑incidence co‑polar transmission T(f) from 211–373 GHz for a single unit cell solved with FEM (periodic boundaries, Floquet ports) and with EMT (staircase discretization of the graded profile and transfer‑matrix propagation). (b) Corresponding co‑polar reflection R(f) over the same band. (c) Band‑averaged transmission $\xoverline{T}_{211-373 GHz}$ versus apex angle $\alpha$ at fixed pitch 
p=0.5mm. (d) Band‑averaged transmission versus pitch p at fixed $\alpha = 22^{\circ}$. The dashed lines indicate the nominal design values used for fabrication of the AR structure.}}
\label{fig_3d}
\end{figure}

\corr{To capture the graded‑index (GRIN) behavior analytically, we also modeled the triangular structure with an effective‑medium theory (EMT) approach \cite{lekner_light_1994, speirs_design_2020}. The groove was divided into N thin layers of thickness $\Delta z$ (staircase approximation). In each layer the local fill factor \textit{f(z)} defined an effective permittivity $\epsilon_{eff}(z)$ using the standard parallel/series mixing rules for E-field parallel and orthogonal to the grooves. The multilayer was then solved with a transfer matrix formulation at normal incidence to obtain the transmission \textit{T(f, $\Delta z$, $\alpha$, \textit{p}, \textit{h})} and reflection \textit{R($\cdot$)} for both polarizations \cite{rytov1956electromagnetic}. The prediction of copolar transmission by EMT was in agreement with the FEM result across 211–373~GHz band (Fig.~\ref{fig_3d}); We then swept the pitch and apex angle around these nominal values in both methods and evaluated the averaged transmission of the band $\xoverline{T}_{211-373 GHz}$. The surfaces of FEM and EMT exhibited the same broad optimum for the apex angle $\alpha$, near the nominal values $\alpha=20^{\circ}$ for $p=0.5~mm$, and predicted a similar mean transmission $\xoverline{T}_{211-373 GHz}$. The broad optimum explains the measured insensitivity to small deviations from the nominal geometry: moderate changes in $\alpha$ and p produce only percent level changes in copolar transmission across the band.}

\corr{Guided by both models and machining constraints, we fixed the AR geometry as follows (see Fig.\ref{schematic}). Each element is an isosceles triangular pyramid in cross section parameterized by pitch p, height h, and angle of apex $\alpha$. We chose p~=~0.5~mm,  h~=~1~mm and $\alpha=22^{\circ}$ which balances fractional bandwidth against robustness to depth and angle errors. The valleys between the pyramids were kept design design and the tips were intentionally truncated to a small flat with width $\leq~0.1~p$ ($\approx$ 50 $\mu$ m in our parts) to reflect the finite mill radius and to ensure repeatable metrology.}

\begin{figure}[!t]
\centering
\includegraphics[width=3in]{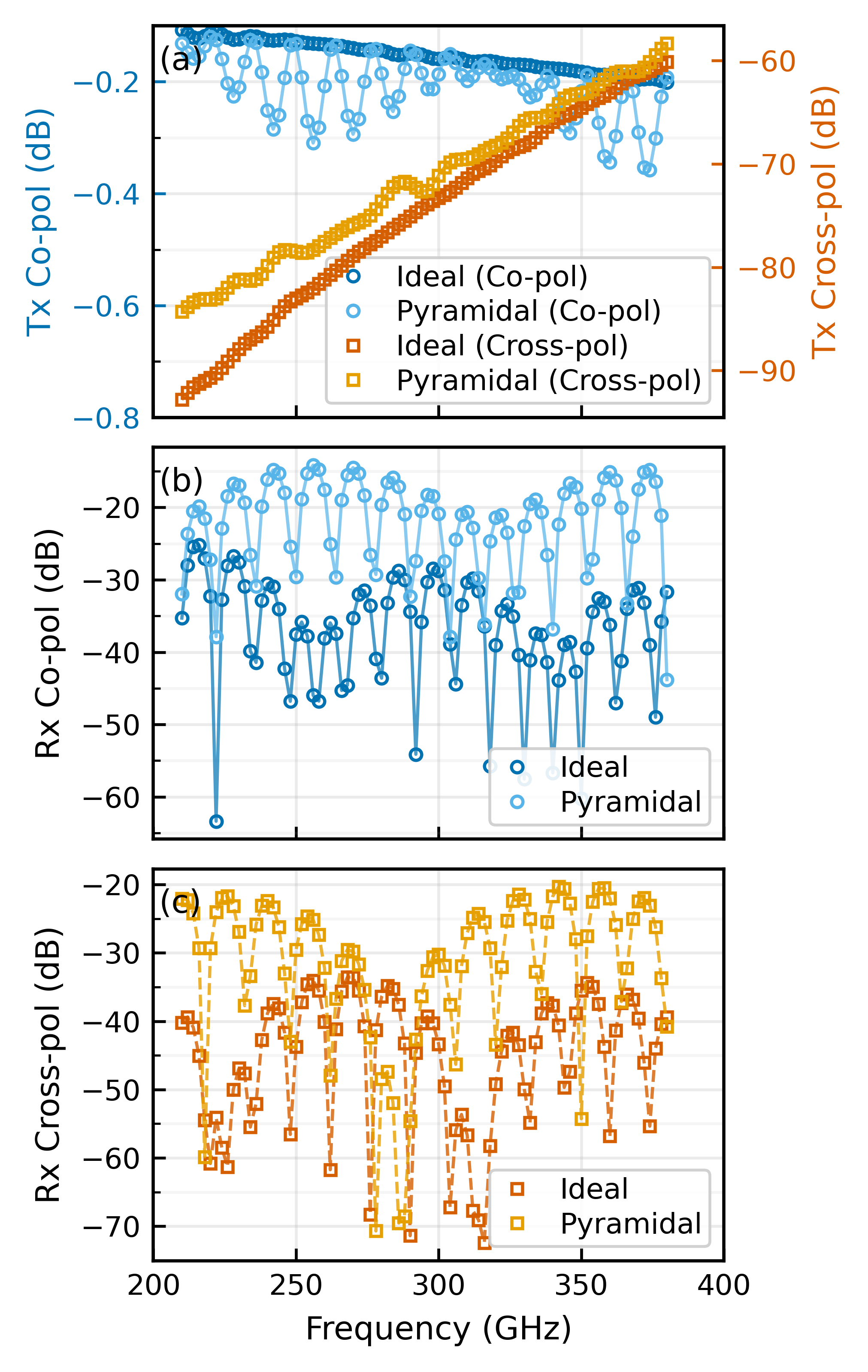}
\caption{Simulated transmission and reflection coefficients of the structured UHMWPE vacuum window at normal incidence, computed via FEM. Two geometries are compared: an idealized profile with sharp pyramidal tips and no flat regions, and the fabricated structure based on experimentally measured groove dimensions, which includes flat-tipped pyramids and inter-groove spacing.
(a) Co-polarized and cross-polarized transmission for incident electric fields aligned with the groove direction on the front surface. Co-polar transmission corresponds to detection with an output port aligned with the incident polarization; cross-polar transmission is measured using a waveguide port rotated by 90°, measuring polarization rotation induced by the structure. (b) Co-polarized reflection.
(c) Cross-polarized reflection. \corr{For all the plots, TE and TM polarisation are perfectly overlaid.}
}
\label{fig_1}
\end{figure}

Fig. \ref{fig_1} compares the simulated co-polar and cross-polar transmission spectra for both the realistic UHMWPE window geometry, which includes the flat-top and valley truncations observed in fabricated samples (see Fig. \ref{profile}), and idealized geometry with perfectly sharp point-tipped triangular grooves without any truncation or valley spacing. The ideal model represents the theoretical upper limit of broadband impedance matching achievable without fabrication constraints, yielding co-polar transmission exceeding 97$\%$ across the entire frequency range of 211–373 GHz, with negligible reflections and a flat spectral response.

Similarly, the realistic model which incorporates flat tips and finite flat regions between pyramids, and machining tolerances, still demonstrates good performance with transmission above 95$\%$ across ALMA Bands 6 and 7. The ripple visible in the co-polar transmission is consistent with weak Fabry–Pérot resonances resulting from imperfect groove terminations, similar to those described in multilayer epoxy coatings \cite{rosen_epoxy-based_2013}. Cross-polarized transmission in both models remains below –60 dB, showing nearly no polarization conversion regardless of geometric imperfections. \corr{ Across 211–373 GHz, the FEM co‑polar transmission and reflection for TE and TM are indistinguishable.} \corr{At normal incidence, the 1D triangular‑groove texture preserves linear polarization, and the co‑polar transmission/reflection amplitudes for E$\parallel$groove and E$\perp$groove are indistinguishable across 211–373 GHz. We observe no measurable polarization‑dependent loss in this band}. \corr{We interpret the result by noting that the back-face grooves are rotated by 90$^{\circ}$, so any small in-plane anisotropy introduced by the front face is compensated by the back face, making the two-sided GRIN window polarization-insensitive, at normal incidence.} We can also conclude that ideal sharp-tipped structures offer a small theoretical gain over realistic CNC-machined geometries which already approaches the fundamental limit of broadband anti-reflection behavior.

\subsection{Fabrication}
CNC machining of UHMWPE’s pyramidal AR grooves must contend with the polymer’s viscoelasticity, relatively low thermal conductivity and high elasticity, which tend to produce built-up edges, chip sticking and workpiece deformation during cutting\cite{piska_advanced_2022}. To address this, we first imported the CAD pattern (pitch p, height h, apex angle $\alpha$) into a Kern precision CNC milling machine and used conical end-mill tool to produce the corrugations with designed period and depth. Because the opposing facet softened and flexed under asymmetric cutting loads, each groove was then hand-finished: a steel blade, with 22$^\circ$ apex angle, was drawn along the deformed side to trim excess material, restore flat-bottom valleys and bring both faces into geometric alignment.

\begin{figure}[!t]
\centering
\includegraphics[width=3in]{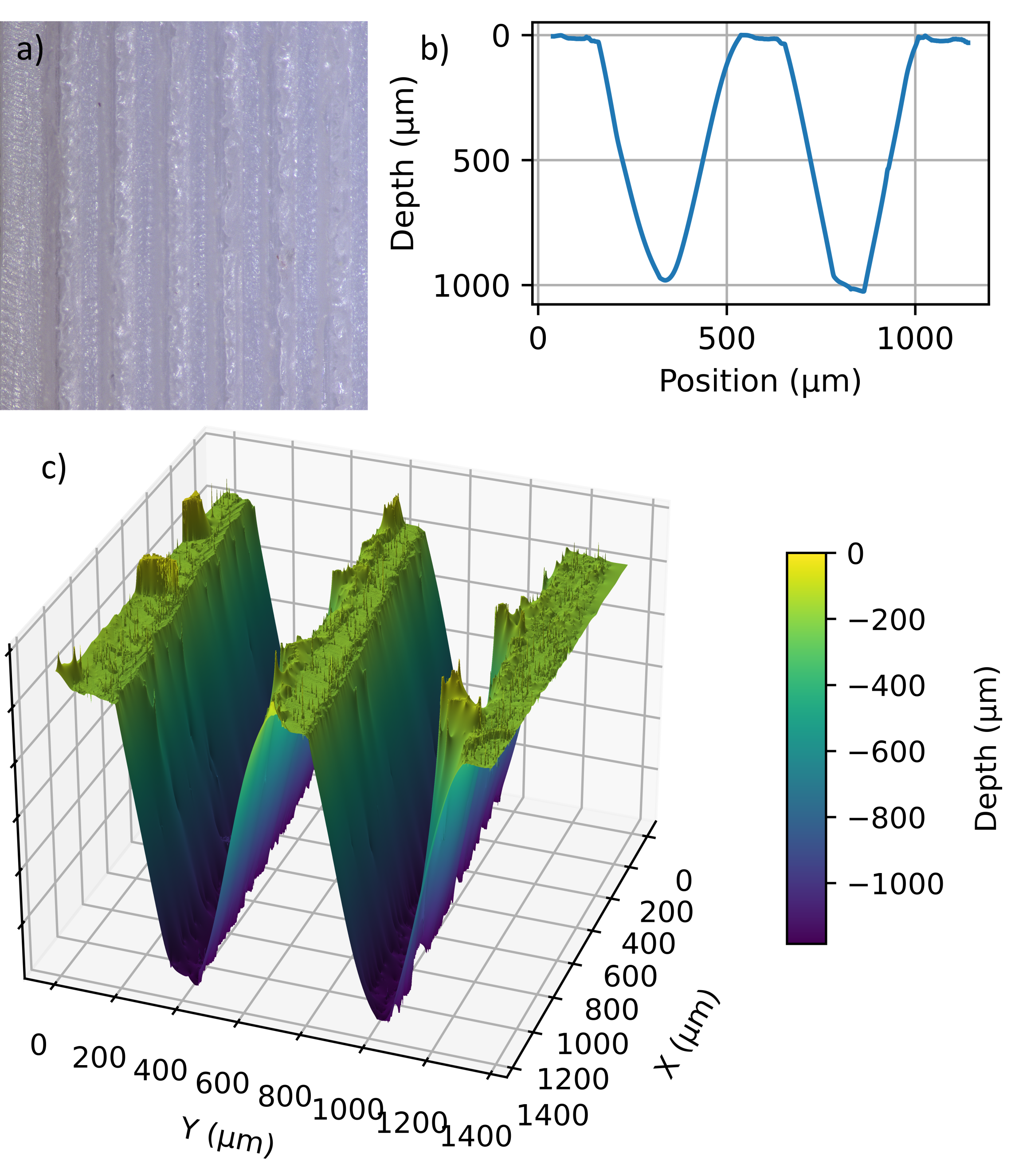}
\caption{a) Top-view optical image of the machined pyramidal corrugation on the UHMWPE window, showing the uniform two-dimensional array of grooves. (b)Optical-profilometer cross-section cut through adjacent grooves, showing a valley flat width of approximately 50 $\mu m$, a truncated pyramid apex flat of 150 $\mu m$, and an apex angle of 15$^\circ$. (c) Three-dimensional optical-profilometer rendering of the corrugation, showing the uniform facet slopes.
}
\label{profile}
\end{figure}

The produced corrugation profile is shown in Fig. \ref{profile}.  The pitch and overall depth of the grooves follow the CAD design with a nominal 50 $\mu m$ flat region between adjacent pyramids. The deformation during milling produced a larger apex flat ($\approx$ 150 $\mu m$) and a shallower apex angle (15$^\circ$ instead of 22$^\circ$). 

\corr{Unless otherwise noted, ‘UHMWPE’ refers to TIVAR$\textregistered$ 1000 Natural Virgin UHMW‑PE.}

\section{Measurements}
\subsection{Time Domain Spectroscopy Measurements}
To characterize the spectral transmission of the vacuum window, we first employed a terahertz time-domain spectroscopy (THz-TDS) system with electro-optic sampling. The system uses a femtosecond pulsed laser to generate and detect broadband THz pulses via photoconductive antennas. The transmitted electric field through the sample is measured in the time domain and Fourier-transformed to yield the complex transmission spectrum \cite{kong_high_2018}. A pair of high-resistivity silicon beam hemispherical \corr{lenses} and a pair of off-axis parabolic mirrors ensure beam collimation and collection. All measurements were performed under dry air purge to minimize absorption by atmospheric water vapor.

The THz beam was linearly polarized, and the samples were oriented such that the grooves on the UHMWPE window were aligned parallel to the electric field. A flat quartz reference window with a single-layer $\lambda/4$ anti-reflection coating in teflon was measured under identical conditions for comparison. The transmission was normalized against a background measurement with no sample to isolate the response of the window structures. Due to small misalignments and residual reflections in the setup, a few data points in the spectrum slightly exceed 100$\%$ transmission. This is an artifact commonly seen in TDS systems and comes from normalisation of the window transmission with vacuum reference. These points are not physically meaningful but are within the expected uncertainty margins.

\begin{figure}[!t]
\centering
\includegraphics[width=3.5in]{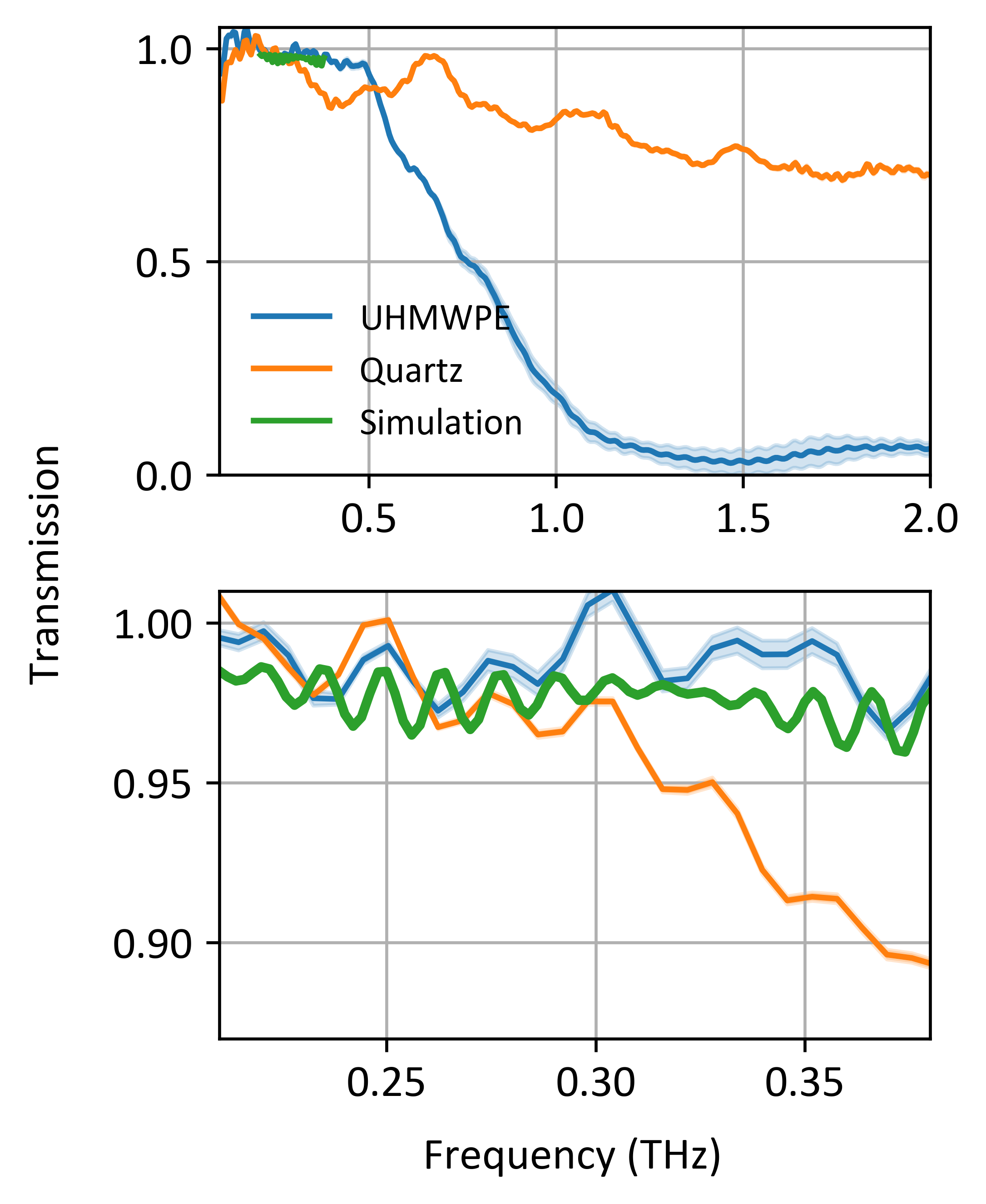}
\caption{Transmission spectra of UHMWPE and quartz windows measured by TDS and comparison with the co-polar simulated response.
Top panel: Full 0.1–2 THz transmission of a 7 mm UHMWPE window with triangular corrugation and a 10 mm quartz window with a single‐layer $\lambda/4$ anti-reflection coating. The UHMWPE exhibits $> 98 \%$ transmission across the band, whereas the AR-coated quartz shows residual ripple from multiple reflections.
Bottom panel: Close-up on the ALMA Bands 6 (211–275 GHz) and 7 (275–373 GHz).
}
\label{fig_2}
\end{figure}

Fig. \ref{fig_2} shows the transmission spectra of the two vacuum window samples. Both windows demonstrate high transmission exceeding 95$\%$ across the full ALMA Band 6 and 7 range. However, the UHMWPE window exhibits superior broadband performance, with transmission between 97$\%$ and 99$\%$ across the entire band and maintaining above 95$\%$ up to 500 GHz. In contrast, the quartz window shows a more narrowband response due to the fixed-index quarter-wave teflon layer. The enhanced performance of the UHMWPE window comes from its gradual impedance transition enabled by the sub-wavelength pyramidal structure.

\subsection{Characterization Using a Heterodyne Receiver}

To complement the broadband spectral measurements obtained with THz-TDS, we further evaluated the vacuum window’s performance under realistic observing conditions using a cryogenic double-sideband (DSB) heterodyne receiver. The basic operation of this receiver is described in \cite{lapkin_wideband_2024}. The receiver covers the 211–373 GHz range of ALMA Bands 6 and 7, and is  equipped with an orthomode transducer (OMT) and an RF hybrid coupler at the input, enabling sensitivity to a single linear polarization (vertical). Achieving a noise temperature between two to four times the quantum noise limit ($T_Q=\hbar \nu/2k_b$) across the band, the receiver provides a highly sensitive platform to measure narrowband transmission and assess any potential impact of the window on receiver noise temperature, impedance matching, or polarization fidelity in a practical system conditions. 

\begin{figure}[!t]
\centering
\includegraphics[width=2.5in]{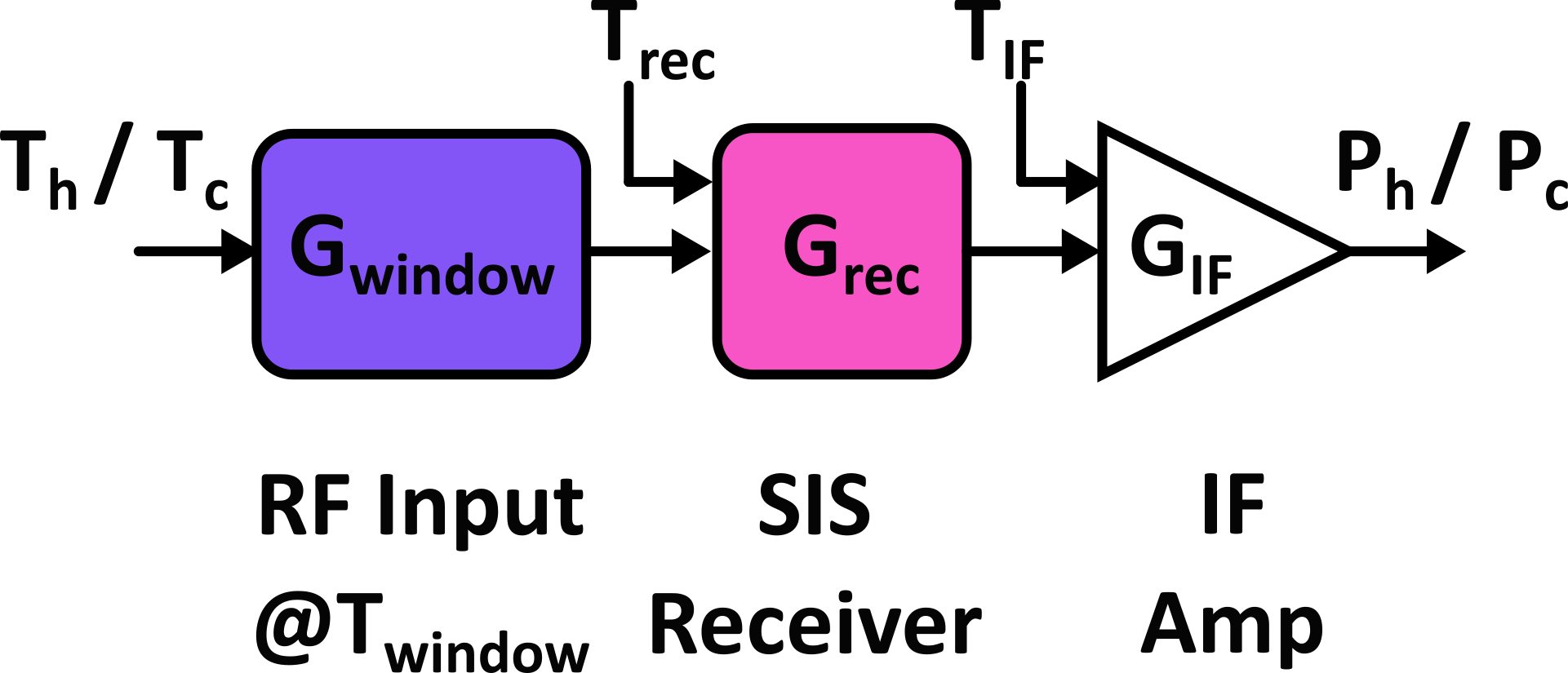}
\caption{Schematic of the broadband SIS receiver with optical losses in front of the mixer
}
\label{Schematic_receiver}
\end{figure}

\begin{figure}[!t]
\centering
\includegraphics[width=3in]{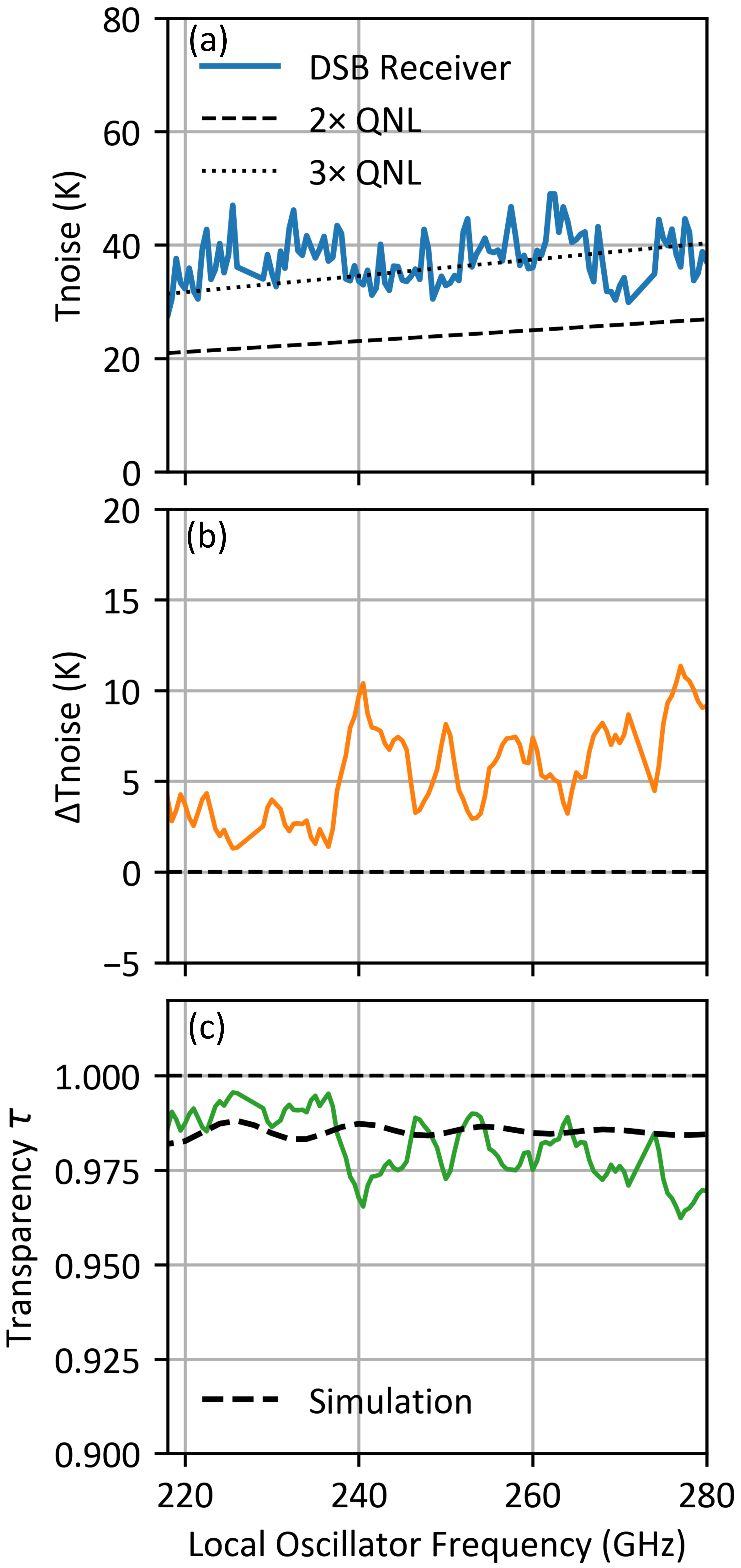}
\caption{DSB receiver performance and window characterization in Band 6 frequency range.
(a) Measured DSB noise temperature as a function of \corr{local oscillator} frequency for the bare receiver, with 2× and 3× quantum‐noise limits overlaid \corr{($\hbar \nu/2k_b$)}. \corr{In these measurements, the receiver was already equipped with a structured UHMWPE infrared filter placed at the 15K stage (see Fig \ref{fig_5})} (b): Change in \corr{DSB} noise temperature ($\Delta T_{noise}$) upon insertion of UHMWPE windows with front‐surface corrugations \corr{oriented perpendicular to the polarisation direction of the receiver}.
(c) Derived optical transparency $\eta$ of the window, computed from the baseline noise difference at $T_{phys}=290K$ along with the simulated transmission.}
\label{fig_3}
\end{figure}

\begin{figure}[!t]
\centering
\includegraphics[width=3in]{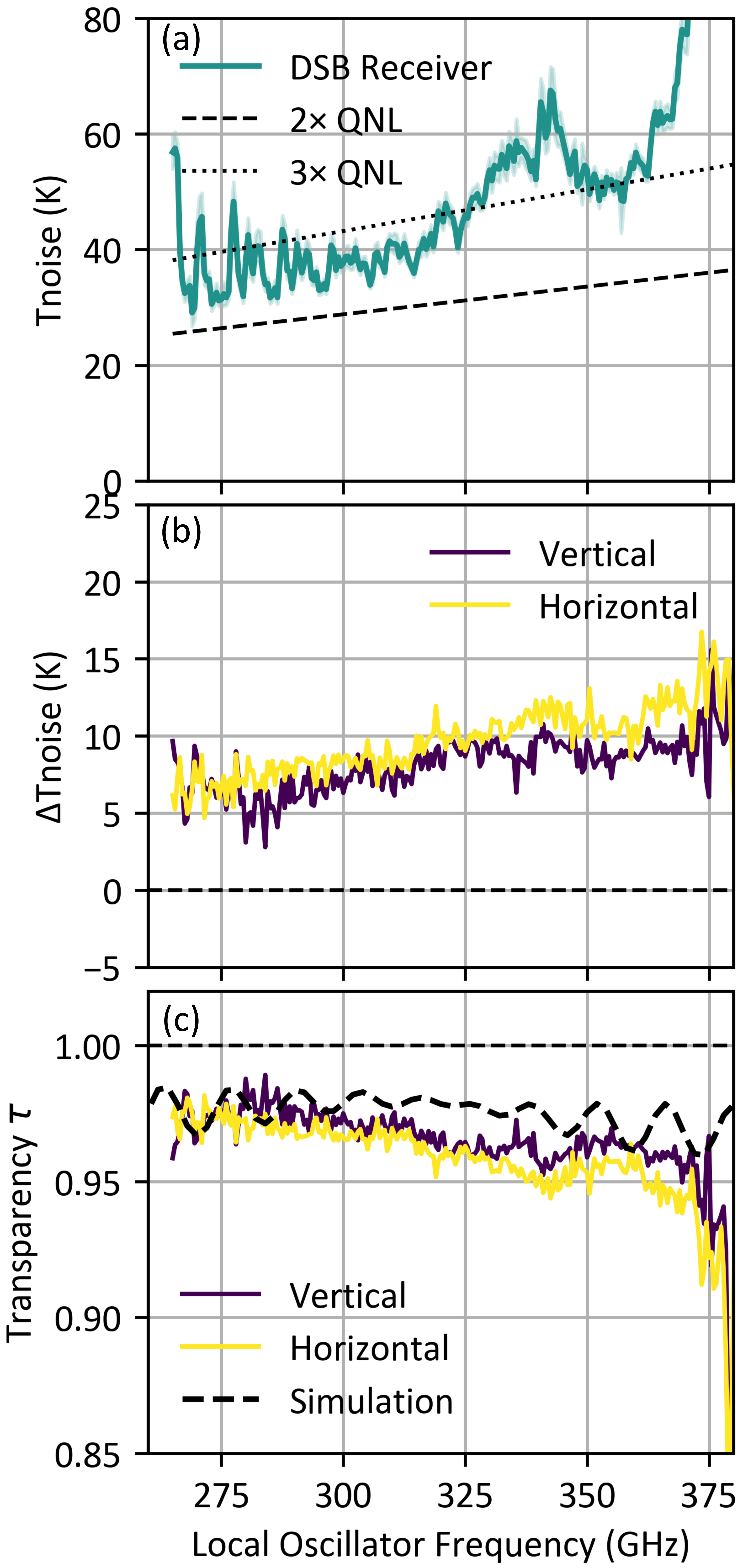}
\caption{DSB receiver performance and window characterization in the Band 7 frequency range. Caption details as in Fig.\ref{fig_3}, but measurements correspond to the higher frequency band (275–373 GHz). Additionally, measurements in (b) and (c) include tests with the window grooves oriented both vertically (aligned) and horizontally (orthogonal) to the receiver’s polarization.
}
\label{fig_4}
\end{figure}

The DSB receiver noise temperature ($T_{rec}$) was characterized separately for ALMA Bands 6 and 7, each using distinct receiver configurations optimized for their respective frequency ranges.  The resulting $T_{rec}$ vs \corr{local oscillator} frequency are presented in Fig. \ref{fig_3}(a) for Band~6 and Fig. \ref{fig_4}(a) for Band~7. To evaluate the impact of the vacuum window, we inserted the UHMWPE sample between the receiver input and a dual blackbody calibration load system consisting of 78~K and 290~K blackbodies. The Y-factor method was used to determine the receiver noise temperature both with and without the window. Notably, the test windows were maintained at 290~K inside a nitrogen-filled chamber to prevent moisture condensation on the samples during measurements. The actual cryostat vacuum windows used in these tests were either a quartz window with Teflon anti-reflection coating specifically made for ALMA Band~6   \cite{vassilev_swedish_2008} or the official ALMA Band~7 windows.

Before calculating the added noise contribution, we first determined the receiver noise temperature $T_{rx}$ in each configuration using standard Y-factor measurements. The Y-factor is defined as the \corr{ratio} of receiver output power for the hot and cold loads,
\begin{equation}
\label{deqn_ex0}
Y=\frac{P_{hot}}{P_{cold}}
\end{equation}
and the corresponding receiver noise temperature is given by:
\begin{equation}
\label{deqn_ex0bis}
T_{rx} = \frac{T_{hot}-YT_{cold}}{Y-1}
\end{equation}
which includes the cumulative contributions from the optics, mixer, and IF amplifier. This establishes a baseline $T_{rx}$ without the test element, against which any subsequent increase can be directly attributed to the insertion loss of the window.

The increased measured receiver noise temperature due to the insertion of the window is shown in Fig \ref{fig_3}(b) and Fig \ref{fig_4}(b). This added noise contribution, $\Delta T$, is then modeled as the result of the window's insertion loss at ambient temperature (290~K). \corr{We treat the window as a passive first stage at physical temperature $T_{w}$ with transmission $\tau$ (loss $L\equiv~1/\tau$). The window has an equivalent input noise $T_{eq, win}~=(L-1)T_w$. We apply Friis:}

\begin{equation}
\label{deqn_ex1}
\corr{T_{rx,~w} =(L-1)T_w+LT_{rx, 0}}
\end{equation}
\corr{where $T_{rx, w}$ and $T_{rx, 0}$ is the receiver noise temperature with and without the window under test, respectively. The excess noise is then:}
\begin{equation}
\label{deqn_ex1}
\corr{\Delta~T\equiv~T_{rx, w}-T_{rx, 0}=(L-1)(T_w+T_{rx,0})}
\end{equation}
so that:
\begin{equation}
\label{deqn_ex2}
\corr{L = 1+\frac{\Delta T}{T_{rx, 0}+T_{w}},~~~~\tau=1/L}
\end{equation}

The derived window transparency lies consistently between 97$\%$ and 99$\%$ across Band~6 and part of Band~7 (Fig \ref{fig_4}(c) and \ref{fig_5}c) and remains above 95$\%$ throughout Band~7 (Fig \ref{fig_4}(c)). Insertion loss corresponds to an added noise temperature in the range of 2–12~K across the band excluding the band edge where the mixer’s DSB configuration begins to capture noise contributions from outside the intended band.

For the measurements in Band~7, we also investigated the depolarization effects introduced by the tested window. The window was inserted in two configurations: first with the grooves oriented vertically, aligned with the receiver’s polarization sensitivity, and second with the grooves rotated by 90~degrees, so that the front-face grooves were orthogonal to the receiver’s polarization. This allowed us to evaluate any polarization-dependent transmission losses or cross-polarization introduced by the structured surface. \corr{Minor differences between the two polarization orientations were observed, which may indicate weak polarization-dependent scattering induced by the groove geometry} (see Fig \ref{fig_4}(c) and (d)), as seen in the simulation (Fig. \ref{fig_1}).

\section{Infrared Filter Insertion Loss Measurements}
To evaluate the performance of the structured UHMWPE as an infrared blocking filter at cryogenic temperatures, we measured the receiver noise temperature in two configurations: first with a conventional Zitex-based IR filter mounted at the 15 K stage, and second with a 2 mm-thick UHMWPE filter featuring the same pyramidal corrugation geometry as the vacuum window.

\begin{figure}[!t]
\centering
\includegraphics[width=3in]{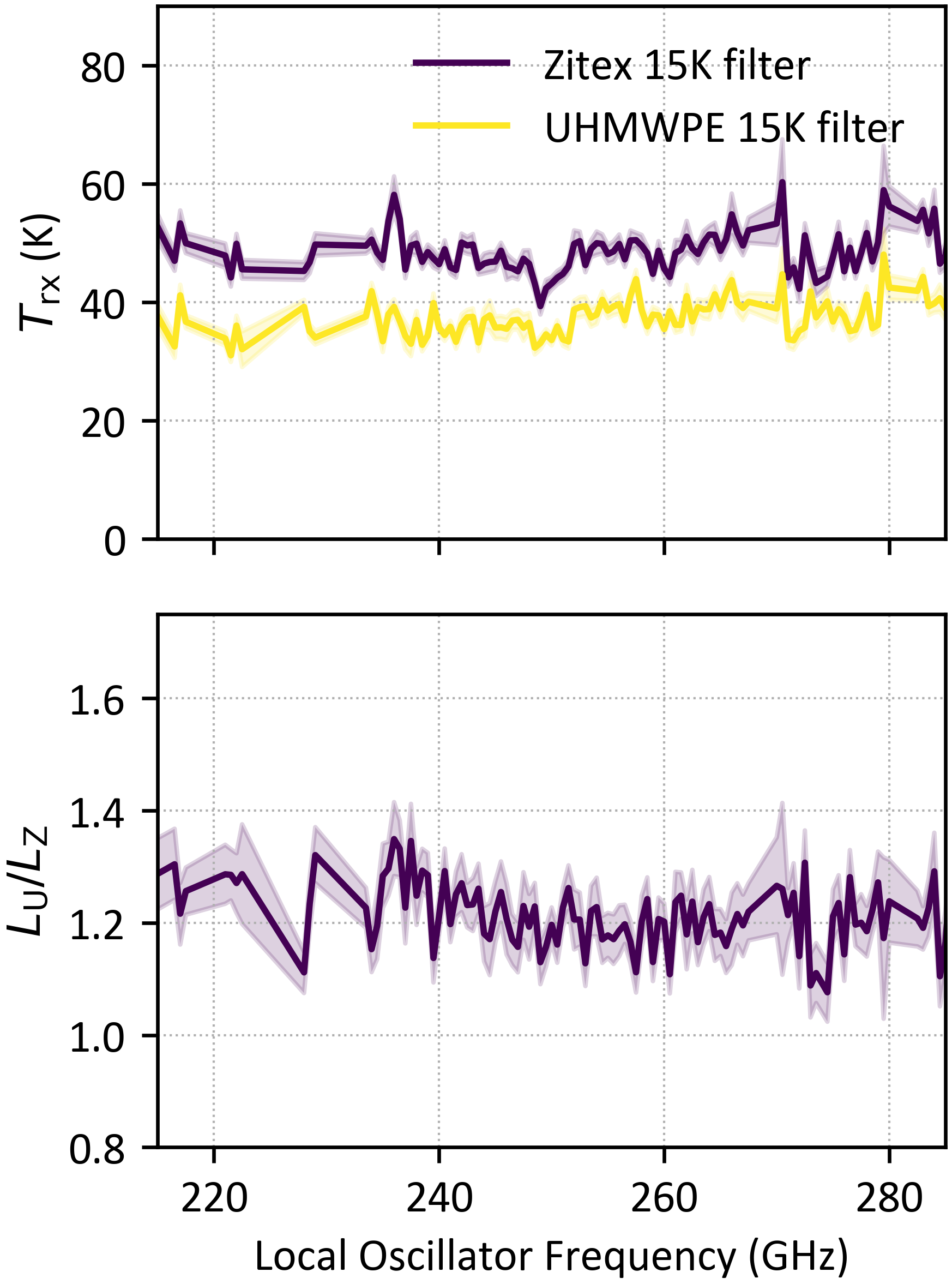}
\caption{a) Measured double‐sideband noise temperature $T_{rx}$ as a function of \corr{the local oscillator} frequency for the heterodyne receiver loaded with a single‐sheet Zitex filter or a UHMWPE filter, both mounted at the 15 K thermal shield. b) Relative insertion loss between the Zitex and UHMWPE filters across the same band.
}
\label{fig_5}
\end{figure}

\corr{Zitex (expanded PTFE) sheets are a long‑standing choice for cryogenic IR blocking in both heterodyne and bolometric receivers because they strongly attenuate mid‑ to far‑IR and add little mm/sub‑mm loss and mass. Measurements show that a thin ($\approx200\mu m$) Zitex sheet transmits $<1\%$ in the 1-50$\mu m$ band but attenuates $\leq10\%$ at wavelengths $\geq200\mu m$, which makes Zitex an effective warm IR blocker/cold mm-wave pass material \cite{benford_optical_2003}.}
\corr{A common implementation is a thermally isolated multi‑layer stack placed between warm optics and the cold receiver. In the idealized limit of
N identical, fully IR absorbing layers with negligible reflection, the transmitted IR power scales as $\frac{1}{N+1}$: one layer halves the load, two layers reduce it to one-third, three to one-quarter, because the layers predominantly emit back toward the warm side rather than into the cold stage. In practice, Zitex layers are not perfectly opaque, but measured single‑layer emissivities and multi‑layer data agree well with this trend \cite{d1999tests}.}
\corr{In contrast, our structured UHMWPE filter is monolithic and mechanically clamped element at the cryogenic shield. The UHMWPE bulk thermalizes to the stage and re-emits absorbed IR at (near) 15~K. Although UHMWPE is not completely opaque at the shortest IR wavelengths, the absorbed power is sunk and re-radiated at the shield temperature instead of a 'floating' intermediate temperature which reduces the net radiative load in the colder stages. High-thermal-conductivity substrates like alumina are also used as cold-anchored absorbers for cryogenic IR filters with negligible thermal gradients across large apertures\cite{inoue2014cryogenic}. Here, our UHMWPE implementation follows the same logic, with pyramidal corrugations providing the broadband AR to keep RF band loss low.}
\corr{We evaluated the 15\,K IR-blocking filters by measuring the receiver noise temperature with Y-factor hot/cold loads for two cases: a conventional Zitex filter (Z) and a 2\,mm-thick structured UHMWPE filter (U) having the same pyramidal-corrugation geometry as the vacuum window. Let $T_{\rm s}=15$~K denote the filter stage temperature, $T_{\rm mix}$ the receiver noise without an IR filter, and $T_{\rm rx,i}$ the receiver noise with filter $i\in\{U,Z\}$ inserted, as derived from the measured Y-factors. For a matched passive element at temperature $T_{\rm s}$ with linear loss $L_i\equiv 1/\tau_i$, Friis’ formula gives:}
\begin{equation}
T_{\rm rx,i} \;=\; L_i\,T_{\rm mix} + (L_i-1)\,T_{\rm s}
\label{eq:friis_irfilter}
\end{equation}
Thus,
\begin{equation}
\frac{L_Z}{L_U} \;=\; \frac{T_{\rm rx,Z}+T_{\rm s}}{T_{\rm rx,U}+T_{\rm s}}
\label{eq:rel_IL}
\end{equation}
\corr{which allows a  comparison of the two filters without knowing $T_{\rm mix}$, the mixer noise temperature without the IR fitlers. These expressions assume the filter is matched at RF so that reflection is negligible compared to absorption, and that the filter is in thermal equilibrium at $T_{\rm s}$.}
\corr{Using the ratio method of Eq.\ref{eq:rel_IL}, we evaluated the relative insertion loss of the Zitex and structured-UHMWPE IR filters at $T_s$=~15~K (Fig.\ref{fig_5}). Across ALMA Band 6. we find $L_Z/L_U>1$, indicating consistenly higher loss for the Zitex filter than for the UHMWPE filter. The band average is $\approx1.20$ corresponding to 0.79~dB more insertion loss for Zitex.}

\corr{The improvement with UHMWPE is consistent with its pyramidal AR texture, which suppresses Fresnel reflections over a wide band, while Zitex is an unstructured expanded‑PTFE sheet with no broadband AR. Additionally, UHMWPE’s higher thermal conductivity aids thermalization of absorbed IR at 15~K, reducing the chance of re‑emission toward the receiver.}

\begin{table}[htbp]
\centering
\caption{Summary of structured UHMWPE vacuum window and IR filter parameters.}
\label{tab_window_summary}
\renewcommand{\arraystretch}{1.3} 
\begin{tabular}{@{} l c @{}}
\toprule
\textbf{Parameter} & \textbf{Value} \\
\midrule
Diameter & 60 mm  \\
Thickness & 7 mm (window), 2 mm (IR filter) \\
Pitch $p$ & 500 $\mu m$ \\
Pyramid height $h$ & 1 mm \\
Apex angle $\alpha$ & 15$^\circ$ \\
Flat region width & 50 $\mu$m (150 $\mu$m) top (bottom) \\
Effective refractive index $n_{\mathrm{eff}}$ & 1.537 \\
Transmission (Bands 6+7) & 95\%–99\% \\
Insertion loss @300K & 2–12 K added $T_{\mathrm{noise}}$ \\
Relative insertion loss (15K IR filter) & $\sim$1.20$\%$ (0.79 dB more loss for Zitex) \\
\bottomrule
\end{tabular}
\end{table}

\section{Conclusion}

We have presented the design, fabrication, and comprehensive characterization of a broadband, structured UHMWPE vacuum window and infrared filter optimized for ALMA Bands 6 and 7. We directly machined pyramidal anti-reflection structures into UHMWPE and achieved broadband impedance matching with simulated and measured co-polar transmission exceeding 97$\%$ across 211–310 GHz, and maintaining above 95$\%$ up to 500 GHz. Heterodyne receiver measurements confirmed minimal added noise temperatures of 2–12 K, validating the low-loss performance of the window under realistic observing conditions. Tests with the window grooves aligned parallel and orthogonal to the receiver polarization showed negligible depolarization effects, which confirms the polarization fidelity. To further improve transmission, the fidelity of the pyramid apex and facet slopes could be enhanced by replacing the end mill with a precision straight or circular blade, allowing both facets to be cut more symmetrically in a single pass.

Furthermore, a 2 mm-thick structured UHMWPE infrared filter at 15 K demonstrated significantly lower insertion loss compared to conventional Zitex filters, with relative losses corresponding to approximately \corr{0.79} dB improvement in average transmission across Band 6. These show that structured UHMWPE is an excellent material for broadband, cryogenically compatible vacuum windows and infrared filters, with superior mechanical strength, low emissivity, and high transmission. The performance makes this technology suitable for next-generation millimeter-wave receiver systems requiring wide bandwidth, low optical losses, and robust cryogenic operation.

\bibliographystyle{IEEEtran}
\bibliography{IEEE_TST}

\end{document}